# Complex impedance of TESs under AC bias using FDM readout system

E. Taralli,[1,a)] P. Khosropanah,[1] L. Gottardi,[1] K. Nagayoshi,[1] M. L. Ridder,[1] M. P. Bruijn[1] and J.R. Gao[1,2]

[1]*SRON Netherlands Institute for Space Research, Sorbonnelaan 2, 3584 CA, Utrecht, The Netherlands*

[2]*Faculty of Applied Science, Delft University of Technology, Delft, The Netherlands*

The next generation of Far-infrared and X-ray space observatories will require detector arrays with thousands of transition edge sensor (TES) pixel. It is extremely important to have a tool that is able to characterize all the pixels and that can give a clear picture of the performance of the devices. In particular, we refer to those aspects that can affect the global energy resolution of the array: logarithmic resistance sensitivity with respect to temperature and current ($\alpha$ and $\beta$ parameters, respectively), uniformity of the TESs and the correct understanding of the detector thermal model. Complex impedance measurement of a TES is the only technique that can give all this information at once, but it has been established only for a single pixel under DC bias.

We have developed a complex impedance measurement method for TESs that are AC biased since we are using a MHz frequency domain multiplexing (FDM) system to readout an array. The FDM readout demands for some modifications to the complex-impedance technique and extra considerations, e.g. how to modulate a small fraction of the bias carrier frequencies in order to get a proper excitation current through the TESs and how to perform an accurate demodulation and recombination of the output signals. Also, it requires careful calibration to remove the presence of parasitic impedances in the entire readout system.

We perform a complete set of AC impedance measurements for different X-ray TES microcalorimeters based on superconducting TiAu bilayers with or without normal metal Au bar structures. We discuss the statistical analysis of the residual between impedance data and fitting model to determine the proper calorimeter thermal model for our detectors. Extracted parameters are used to improve our understanding of the differences and capabilities among the detectors and additionally the quality of the array. Moreover, we use the results to compare the calculated noise spectra with the measured data.

**I. INTRODUCTION**

Transition edge sensor (TES) microcalorimeters [1] are very versatile superconducting devices, which can be used to detect radiation in a wide energy range e.g. from γ-ray down to submillimeter [2-8]. A TES consists of a superconducting thin film, typically with a transition temperature $T_c$~100 mK, which is strongly coupled to its absorber but weakly thermal coupled to a lower temperature heat bath via a thermal conductance $G$. In principle, TESs operate as thermometers: the absorption of incident photons by means of the absorber heats the device, which is biased in the transition between the superconducting and the normal states, causing a change in the resistance that is proportional to the photon energy absorbed. This variation is read out using a superconducting quantum interference device (SQUID).

---

[a)] Electronic mail: e.taralli@sron.nl.

Large TES arrays are under development by many groups worldwide for ground-based and space-based applications [9-12] and different technologies to readout a large number of detectors are also under development [13-17]. At the Netherlands Institute for Space Research (SRON), we are currently pursuing high-performance arrays of TESs [11] together with a MHz frequency domain multiplexing (FDM) readout system demonstration, where each SQUID channel reads 40 pixels out [18]. Note that the combination of an large TES array and FDM readout system becomes the key technology in most of the next generation of space observatories [19-22].

Our X-ray TESs with different aspect ratios are based on a superconducting TiAu bilayer on a silicon nitride (SiN) membrane and coupled with different size Au or Au/Bi absorbers. Other variations include TESs with or without Au bars, with or without slots in the membrane and a variety of absorber-TES couplings. Our typical detectors show a full-width-at-half-maximum (FWHM) energy resolution $\Delta E_{FWHM}$ < 4 eV at 6 keV [11].

The FDM [17, 23] is used to read out a TES detector array. It applies a set of sinusoidal AC carriers, which bias the TES detectors at their working points and are amplitude modulated when the TES detectors are hit by X-ray photons. The detectors are separated in frequency by placing them in series with LC resonators, each having a specific frequency. The frequency bands assigned to the detectors are separated to prevent the detectors from interacting with each another. This allows the readout of multiple TES pixels by one amplifier channel, which uses only one set of SQUID current amplifiers. We are currently using an 18-channel FDM readout system with 1-5 MHz bias frequencies, which is a small version of the baseline readout based on the 40-channel FDM readout demonstrated in Ref. [22].

In this scenario, it is extremely important to have a tool that is able to get as much information as possible from the detectors placed on an array. Thermal and electrical parameters like logarithmic resistance sensitivity with respect to temperature and current ($\alpha$ and $\beta$ parameters, respectively), uniformity of TESs and the correct understanding of the calorimeter thermal model play an important role in the evaluation of the global energy resolution of the array. The complex impedance measurement of a TES [24] is a powerful technique that is able to give all these information at once, which is well-established under DC bias [24-27] but also modified to be used for single pixels under AC bias at a relatively low bias frequency (~400 kHz) by adding white noise as a small signal excitation and compared that with DC bias results [28-30].

There is also a technique, developed under FDM that focuses on the "complex thermal conductance" of the TESs that are slowed down by adding extra heat capacity to ensure the stability of the readout system [31]. In this technique the thermal response of a detector as a function of frequency is probed by varying the frequency of an added excitation tone near the carrier frequency but the report comes short on extending the findings to the complex impedance parameters and expected detector noise.

In this work we describe the details of a method for measuring the complex impedance of TESs that are biased in AC using a Base Band Feed-Back (BBFB) FDM readout system. By means of this technique we are able to measure and characterize easily all the pixels of an array, giving a clear picture of the performance of the detectors. In particular in Sec. II, after a brief description of the uniform 5×5 TESs array under test, we explain in the detail the complex impedance measurement technique under AC bias. We discuss the main challenges of this technique e.g. how to modulate a small fraction of the bias carrier frequencies in order to get a proper excitation current through the TESs and how to perform an accurate demodulation and recombination of the output signals. Measurement calibration is also presented because all the parasitic impedances have to be taken into account. In Sec. III we show a complete set of complex impedance measurements for some detectors located in the array. We discuss the statistical analysis of the residuals between the model fitting presented in Sec. II and the impedance data to determinate the validity of the calorimeter thermal model used to describe our detectors. Noise spectra of the detectors at specific bias points are also reported. In Sec. IV we discuss the results, showing the common features among TESs of the same type i.e. with or without normal metal Au bar structures and at the same time highlighting the main differences between these two types of TESs. Specifically, we examine $\alpha$ and $\beta$ behaviors, uniformity of the array and theoretical noise model comparison.

## II. TES DETECTORS AND EXPERIMENTAL SETUP

### A. Detectors

Six out of 25 TESs in a uniform 5×5 array were wired for AC complex impedance measurements (see Fig. 1 on the left). Vertical slots in the SiN membrane realize the thermal isolation between the devices in each row. Some of the devices have a relatively higher thermal conductance ($G$~350 pW/K) and some others have a lower value ($G$~150 pW/K), which is realized by having horizontal slots in the SiN membrane in addition to the vertical ones. Eventually we connected four TESs with bars and two TESs without bars. All the TESs are placed on a 1-μm thick SiN membrane and have the same size of 140×100 μm$^2$ and the same bilayer thickness of Ti (20 nm) and Au (50 nm), with $T_c$ ~ 100 mK and normal resistance $R_N$ ~ 220 mΩ. All the absorbers also have the same size (248×248 μm$^2$) and the same thickness (3 μm of Au and 3.5 μm of Bi). Each absorber has four contact points to the membrane at the corners (see Fig.1 on the right) but some of them have an additional contact point in the center of the TES. More details on the TES array can be found in [11].

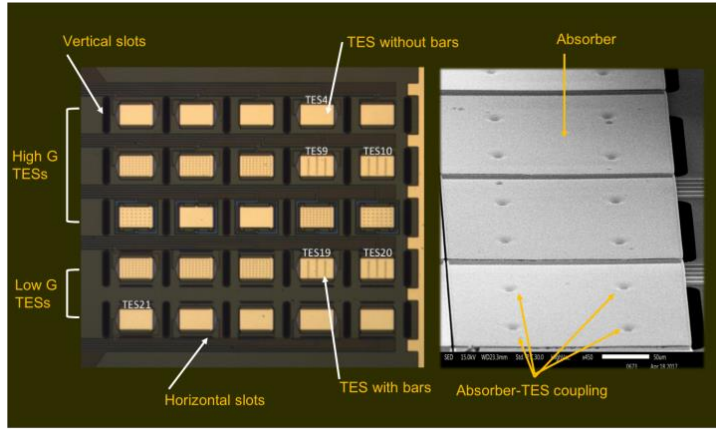

FIG. 1. Photo of the TES array under test before fabrication of the absorbers (left). The first three rows are high $G$ devices and the last two rows are low $G$ with additional horizontal slots at the top and the bottom of the TESs. A total of six TESs are characterized and are numbered as 4, 9, 10, 19, 20 and 21. Example of absorber and absorber-TES coupling (right). Coupling can be done by means of pillars at the four edges of the absorber or with five pillars at the four corners and one in the center.

### B. FDM readout and complex impedance

In our FDM readout system a TES is biased by a carrier signal $B_c \cos \omega_c t$ with $\omega_c = 2\pi f_c$ and $f_c$ between 1 and 5 MHz, generated by digital electronics with 20 MHz sampling rate. The different bias frequencies $f_c$ are defined by a high-Q superconducting LC resonator filter chip, similar to the one reported in [32]. This consists of 18 resonators separated in frequency by 200 kHz with a coil inductance $L = 400$ nH and a capacitance ratio $C/C_b = 9$, where $C$ is the main capacitance of the resonator and $C_b$ is the bias capacitance. The TES current is picked up by a two-stage SQUID assembly, consisting of a low-power single SQUID at the mK stage and a high-power SQUID array at the 2K stage. This provides the pre-amplification of the summed signals to a level sufficiently above electromagnetic interference (EMI) and electromagnetic compatibility (EMC) noise sources so that the dynamic range of the readout chain is not reduced. The signal is further amplified by a low-noise amplifier (LNA) at room temperature and digitized. The carrier of this signal has the same frequency as the bias but different amplitude and phase due to the bias and readout circuit transfer function and can be written as: $A_c \cos(\omega_c t + \theta)$. This signal is then demodulated using the original carrier, resulting in the quadrature $I$ and $Q$ signals that are proportional to $\cos \theta$ and $\sin \theta$, respectively. The carrier can be phase shifted to change $\theta$ (carrier rotation) before it is used for demodulation. Typically, we use the carrier rotation to set $\theta = 0$ when the TES is in its normal state and purely resistive in order to have the entire demodulated signal in $I$, while $Q$ is kept close to zero. When the TES is in the transition, $\theta$ varies as a function of the bias point, which is a signature of the weak-link phenomenon [33]. However, the Josephson current measured in the TiAu TES microcalorimeters developed at SRON, with high power $P$ and high normal resistance $R_N$, is very small [34].

We use the FDM readout in a closed-loop [23], where the $I$ and $Q$ signals are re-modulated and combined to provide a feed-back to the SQUID in order to enhance the dynamic range and the linearity. Before the demodulated signals are acquired, they go through a decimation filter that can be adjusted for the desired down-sampling. In this case we use a decimation factor of 128 which reduces the sampling rate to 156.25 kHz (= 20 MHz/128).

The general idea of the complex impedance measurement is to bias a TES in the transition and to measure the response of the detector (in amplitude and phase) to a small signal AC excitation, added to that bias voltage. Doing so for many bias points in the transition and for many excitation frequencies within the detector bandwidth, gives a thorough picture of the essential detector thermal and electrical parameters.

In our FDM system, the only way to add a small AC signal excitation to the bias line in the detector band is to modulate a small fraction of the bias carrier $b_m \ll 1$ with a cosine wave at a low frequency $\omega_m = 2\pi f_m$. The TES bias can be written as:

$$(1 + b_m \cos \omega_m t) B_c \cos \omega_c t =$$
$$B_c \cos \omega_c t + 0.5\, B_c b_m [\cos(\omega_c + \omega_m)t + \cos(\omega_c - \omega_m)t] \qquad (1),$$

which is equivalent to applying upper-side band and lower-side band excitations at the same time. The signal that is picked up by the SQUID has this form: $(1 + a_m \cos(\omega_m t + \varphi)) A_c \cos(\omega_c t + \theta)$ and the corresponding $I$ and $Q$ signals are:

$$I \propto \cos \theta \ (1 + a_m \cos(\omega_m t - \varphi)) \qquad (2),$$
$$Q \propto \sin \theta \ (1 + a_m \cos(\omega_m t + \varphi)) \qquad (3).$$

The data acquisition is triggered by the small AC signal excitation that modulates the bias ($\cos \omega_m t$) to maintain the same phase condition of the input signal through out the measurement. Note that due to the decimation filter, the sampling rates for the small AC signal and the data acquisition are different (i.e. 20 MHz and 156.25 kHz, respectively). This means that triggering only works perfectly at specific frequencies where $f_m = 156250/2^n$, with $n$ being an integer. Slight jitters occur at other frequencies, which are removed by averaging. At each frequency (110 different $f_m$) the demodulated current is measured ten times as a time series, with 65536 samples (0.42 seconds long) each and averaged. The result is then fitted with a cosine function to extract the amplitude and the phase. Dividing the input voltage by the output current gives us the measured complex impedance $Z_m$.

In principle both $I$ and $Q$ signals hold the amplitude and phase information of the TES response and either or both can be used for analysis. We use the carrier rotation function to set $\theta = 0$ in the normal state so that $I \gg Q$. Therefore, the cosine quadrature $I$ signal is typically used for our impedance analysis. In our measurement $b_m = 0.01$ (1% of the carrier amplitude) and $f_m$ varies from 5 Hz up to 10 kHz. Note that an excitation that is too small obviously leads to a noisy measurement and too large an excitation can modify the selected bias point and moreover induces nonlinear effects generating unwanted higher harmonics of the tone.

In order to extract the impedance of the TES, the effect of the bias and readout circuits on the measured impedance needs to be calibrated out. The measured impedance $Z_m$ can be written as:

$$Z_m = (Z_{TES} + Z_{bias})\, T \tag{4},$$

where $Z_{TES}$ is the impedance of the TES, $Z_{bias}$ is the Thévenin equivalent impedance of the voltage bias circuit and $T$ is the transfer function of the current readout circuit. Since we know that the impedance of the TES in superconducting state is zero and its impedance in the normal state is $R_N$, the measured impedance in these two states can be written as:

$$Z_m^S = Z_{bias}\, T \tag{5}$$

$$Z_m^N = (R_N + Z_{bias})\, T \tag{6}.$$

Knowing the $R_N$ and measuring the $Z_m^N$ and $Z_m^S$, we can solve the above equations for $Z_{bias}$ and $T$:

$$Z_{bias} = R_N \frac{Z_m^S}{Z_m^N - Z_m^S} \tag{7}$$

$$T = \frac{Z_m^N - Z_m^S}{R_N} \tag{8}$$

Finally, inserting (7) and (8) in (4), the TES impedance at a specific bias point in the transition can be extracted from the measured data as:

$$Z_{TES} = R_N \frac{Z_m - Z_m^S}{Z_m^N - Z_m^S}. \tag{9}.$$

Obviously, it is important to make sure that the TES is completely in the normal state when measuring $Z_m^N$. It could happen that the TES resistance has an observable slope in resistance above $T_c$ and it still has some responsivity at those bias points that appear to lie in the normal part in the IV curves. Using the impedance data measured in these points as $Z_m^N$ for calibration results in a faulty $Z_{TES}$ set. To avoid this, it is advisable also to measure the impedance when the TES is thermally normal. The impedance $Z_{TES}$ is then fitted with the following three free parameters as:

$$Z_{TES} = Z_\infty + (Z_\infty - Z_0)\frac{1}{-1 + i\omega \tau_{eff}}, \tag{10},$$

where $Z_0$ is the low-frequency limit of the impedance, $Z_\infty$ is the high-frequency limit of the impedance and $\tau_{eff}$ is the effective time constant of the detector. From these three parameters we can derive $\beta = \partial \ln R / \partial \ln I$, the loop gain of the electro-thermal feedback at low frequency $\mathcal{L}$, the heat capacity $C$ and $\alpha = \partial \ln R / \partial \ln T$ as follows [1]:

$$\beta = \frac{Z_\infty}{R} - 1 \tag{11}$$

$$\mathcal{L} = \frac{Z_0 - R(1+\beta)}{Z_0 + R} \tag{12}$$

$$C = \tau_{eff}\, G\, (1 + \mathcal{L}) \tag{13}$$

$$\alpha = \frac{\mathcal{L}GT}{P},  \quad (14),$$

where $R$ is the TES resistance at the specific bias point, $G$ is the thermal conductance derived from the P(T) curve, $T$ is the TES temperature and $P$ is the Joule heat dissipated in the TES at that bias point.

Note that $Z_0$ is a negative number with an absolute value very close to $R$. This means that if the measurement is affected by noise at low frequency, the denominator in (12) could falsely turn into a positive number, leading to a negative value for $\mathcal{L}$ and consequently a negative number for $C$. Since the heat capacity of the absorber (~1.18 pJ/K) is two orders of magnitude larger than the heat capacity of the TES (~0.02 pJ/K) we can consider it to be constant over all the bias points and neglect the variation that occurs during the phase transition. This consideration follows the assumption that our absorber is strongly coupled to our TES, leading us to use the one-body model for the fitting analysis as mentioned previously. In this way, $\mathcal{L}$ can be calculated from (13) instead of (12), avoiding the effects of low-frequency noise.

### III. RESULTS

We started characterizing four TESs with bars: TES19-TES20 with relatively low $G$ values of 156 and 132 pW/K and TES9-TES10 with higher $G$ values of 300 and 311 pW/K, respectively. These TESs were biased with carrier frequencies of 1.6, 3.4, 3.8 and 3.2 MHz, respectively. During the measurement on a specific TES, all the other detectors were kept unbiased to avoid any detector-to-detector cross talk due to current carrier leakage between neighbour channels. Fig. 2 shows as an example the measured $Z_{TES}$ of TES9 calibrated using (9) for four bias points (dots) and the corresponding fit (lines) using the simple one-body model as in (10). We fit the real part $Re_{Z_{TES}}(f)$ and the imaginary part $Im_{Z_{TES}}(f)$ of the TES impedance at the same time at each bias point. We observe that the one-body model can explain well all the devices presented in this paper. The quality of the fit is quantified by looking at the histograms of the residuals in the real and imaginary parts. Fig. 3 shows this analysis for TES9 as an example and fits with similar quality were obtained for all the others TESs. In the upper plot (a) we have the histogram of the residuals for the real part or in other words the discrepancy between the observation (real part of the experimental data) and the expectation (real part of the one-body model) for each frequency and for every bias point; the histogram in the right plot (b) has the same meaning as the previous one, but describes the residuals of the imaginary part of the impedance. We can already note that both of the histograms of residuals approximate the Gaussian distribution or random errors, making the relationship between the explanatory variables and the predicted variable a statistical relationship. Therefore, the fact that the residuals appear to behave randomly suggests that the model fits the data correctly. The center plot is the conjunction of the two histograms, which shows that most of the residuals are concentrated around zero. We would like to stress that we are looking at the residuals evaluated for each frequency both in the real and imaginary part and for every bias point (27 in this specific case) giving a total of 4600 residuals, where 4483 out of those (~97%) are included in the graph. In Fig 4. we also report the total residuals for all bias points both for the real and imaginary parts obtained from the data analysis using the one-body model compared with the corresponding residuals obtained from the two-body model. The lower standard deviation of the histogram on the left (a) compared with the one on the right plot (b) indicates again that the one-body model seems to be good enough to explain our detectors and experimental data.

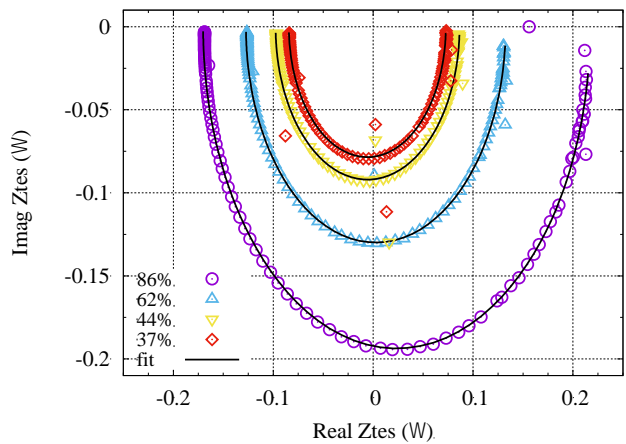

FIG. 2. Measured impedance $Z_{TES}$ for TES9 as an example of detectors with bars. It shows four different bias points (dots) and frequency between 5 Hz and 10 kHz and the corresponding fit with the linear model (lines) to the impedance data with three free parameters $\alpha$, $\beta$ and $\tau_{eff}$.

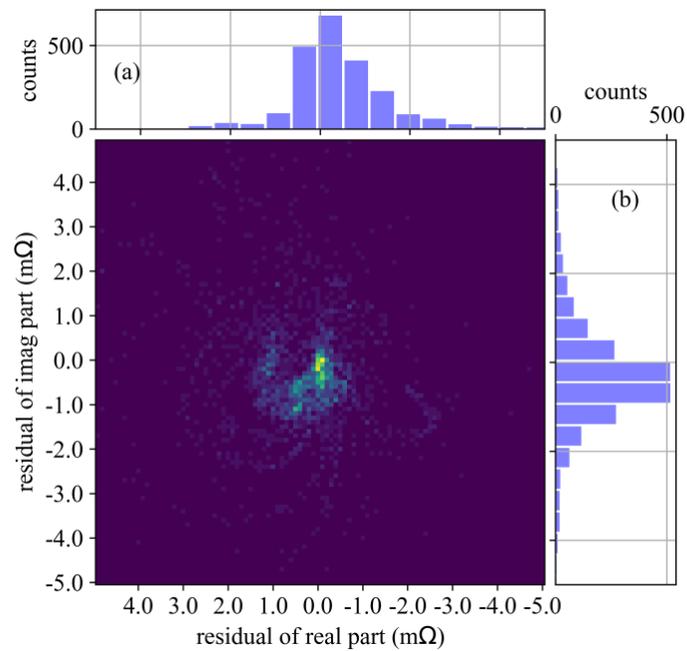

FIG. 3. Plot (a) and (b) are the histograms of the residuals between the experimental data and the fitting model for each frequency and for every bias point related to the real and imaginary part of the impedance, respectively. The plot in the center is the conjunction of the two histograms and it shows qualitatively how many residuals are placed around the origin. For more explication see the text.

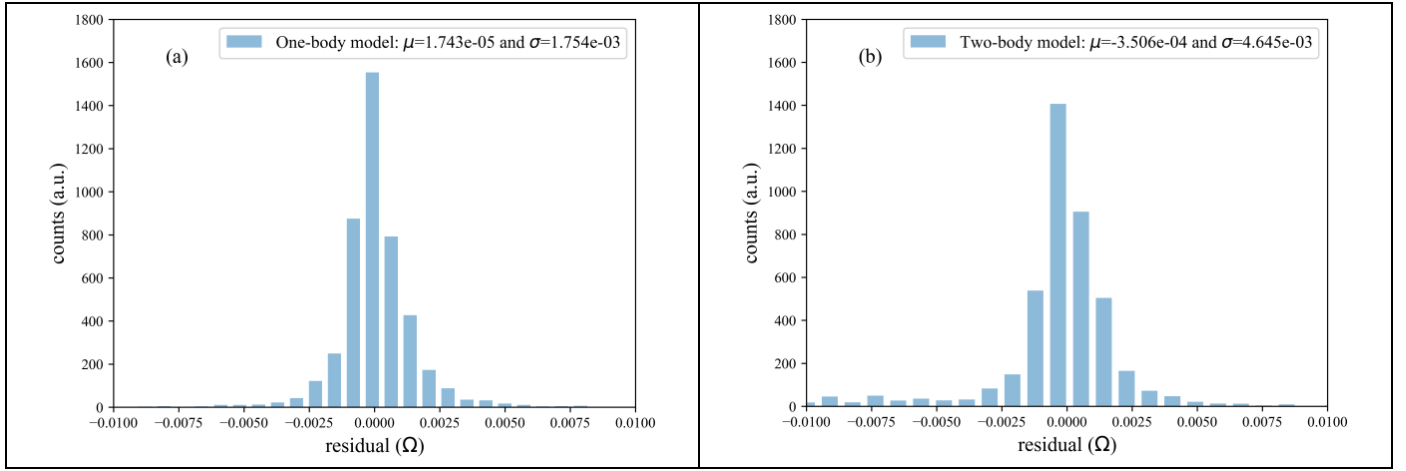

FIG. 4. Histograms of the total residuals for all bias points both for the real and imaginary part obtained by using (a) one-body model and (b) two-body model.

We obtained values of $\alpha$, $\beta$, $\tau_{eff}$ and $\mathcal{L}$ fitting the complex impedance measurements for the TESs with bars and these are plotted as a function of TES resistance in Fig. 5.

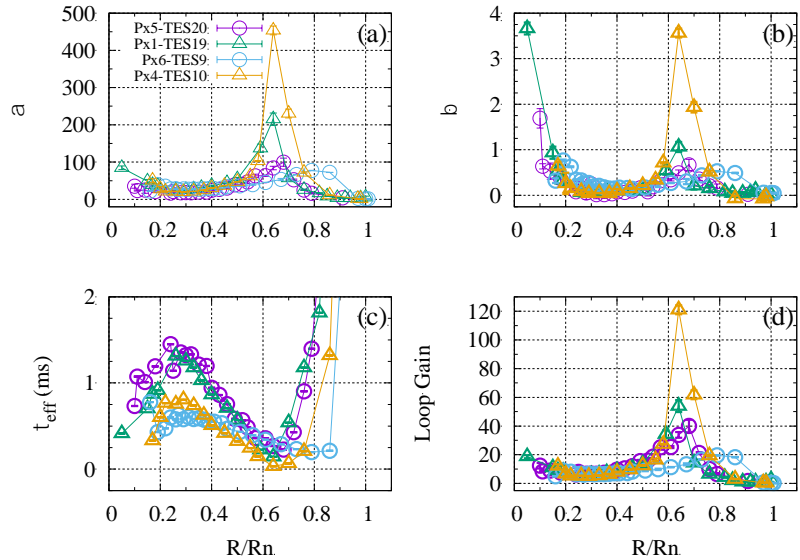

FIG. 5. Parameters derived from impedance measurements: $\alpha$ (a), $\beta$ (b), $\tau_{eff}$ (c) and $\mathcal{L}$ (d) for TESs with bars over the measured bias points. Values of $\tau_{eff}$ corresponding to bias points $R/R_N$ higher than 0.9 are intentionally left off-scale in plot $c$. Errors have been propagated but the corresponding error bars are too small to be properly appreciated from the plots. Lines serve no other purpose than to guide the eye.

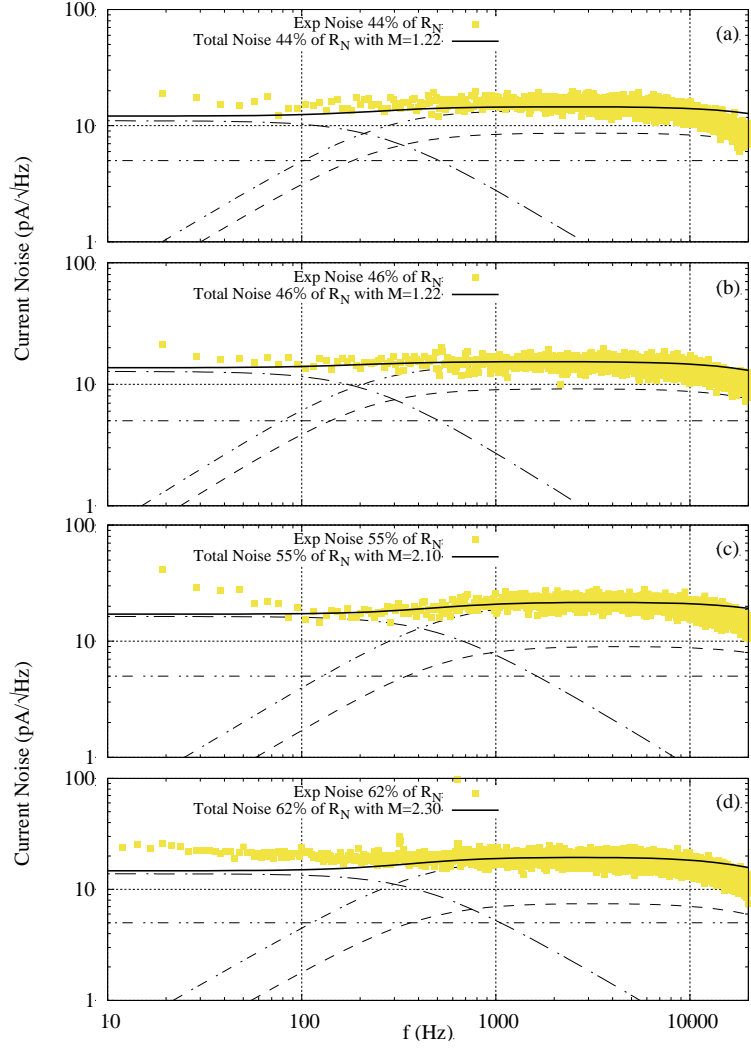

FIG. 6. Noise measurements of TES20 (a) at 44% of $R_N$, TES19 (b) at 46% of $R_N$, TES10 (c) at 55% of $R_N$ and TES9 (d) at 62% of $R_N$. Noise contributions: SQUID noise (dot-dot-dash line), Johnson noise (dashed line), Excess Johnson noise (dot-dash line) and phonon noise (dot-long dash line). The discrepancy between the measured and calculated noise at frequencies above 10 kHz is due to the use of a band-pass filter to avoid interference with the neighboring pixel.

One can use the parameters obtained from the complex impedance to model the detector noise. In Fig. 6 the measured noise spectra are shown for all four TESs: TES20 (a) at 44% of $R_N$, TES19 (b) at 46% of $R_N$, TES10 (c) at 55% of $R_N$ and TES9 (d) at 62% of $R_N$ and the results from the model are over-plotted. The model noise contributions are: phonon noise, TES Johnson noise, excess Johnson noise and SQUID noise. Those noise sources describe very well the noise observed at frequencies higher than 100 Hz, while at lower frequencies we observe a discrepancy mainly due to the effect of the pulse tube cooler. The discrepancy between the measured and calculated noise at frequencies above 10 kHz is due to the use of a band-pass filter to avoid interference with the neighboring pixel. In the frequency range where the Johnson noise is dominant there is an excess noise, which is quantified as $M$ times the Johnson noise and introduced by this factor $M$ [2].

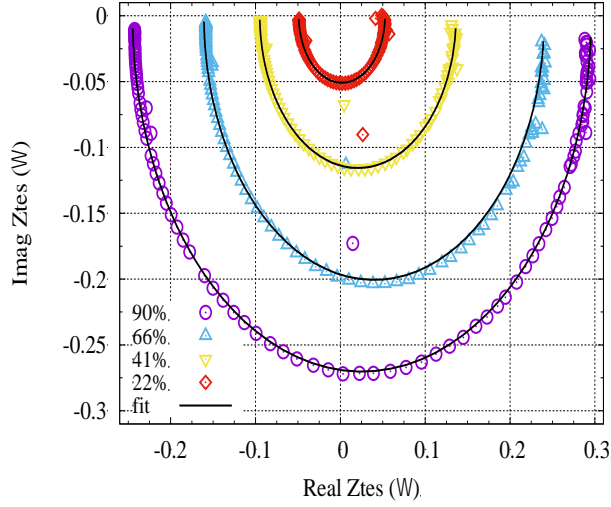

FIG. 7. Measured impedance $Z_{TES}$ for TES21 as an example of detectors without bars. The plot shows four different bias points (dots) between 5 Hz and 10 kHz and the corresponding fit with the linear model (lines) to the impedance data with three free parameters $\alpha$, $\beta$ and $\tau_{eff}$.

We also characterized two TESs without bars: TES4 with high $G$=339 pW/K and TES21 with low $G$=140 pW/K biased at carrier frequencies of 2.9 and 4 MHz, respectively. Fig. 7 shows the measured impedance $Z_{TES}$ for TES21 at four different bias points. We obtained the values of $\alpha$, $\beta$, $\tau_{eff}$ and $\mathcal{L}$ for both TESs as a function of TES resistance as shown in Fig. 8.

We measured the noise spectra for both TESs and in Fig. 9 we show the noise measurement for two specific bias points and the corresponding theoretical model using the parameters obtained from the fitting of the complex impedance curves.

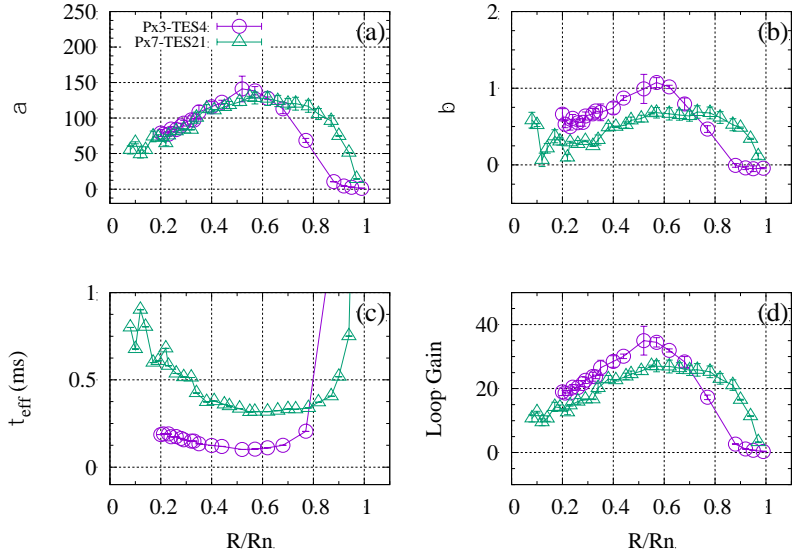

FIG. 8. Parameters derived from impedance measurements: $\alpha$ (a), $\beta$ (b), $\tau_{eff}$ (c) and $\mathcal{L}$ (d) for TESs without bars at all the measured bias points. Values of $\tau_{eff}$ corresponding to bias points higher than 0.8 are intentionally left off-scale in plot $c$. Errors have been propagated but the corresponding error bars are too small to be properly appreciated from the plots. Lines serve no other purpose than to guide the eye.

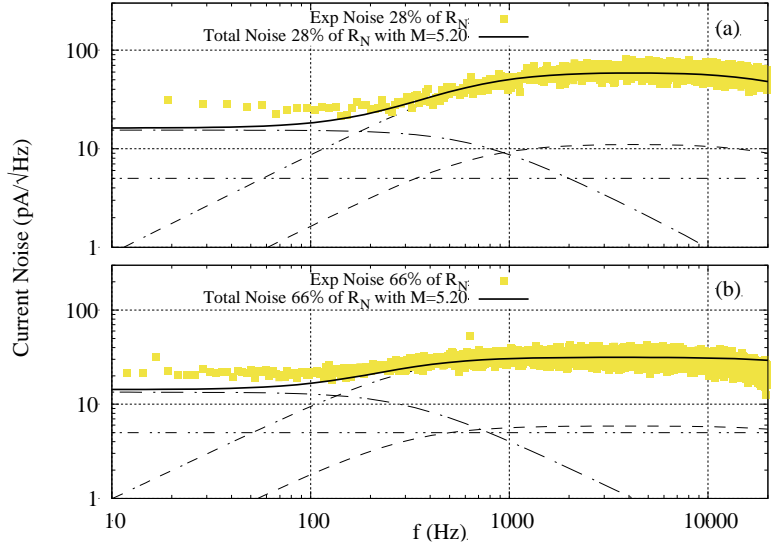

FIG. 9. Noise measurements of TES4 (a) at 28% of $R_N$ and TES21 (b) at 66% of $R_N$. Noise contributions: SQUID noise (dot-dot-dash line), Johnson noise (dashed line), Excess Johnson noise (dot-dash line) and phonon noise (dot-long dash line). The discrepancy between the measured and calculated noise at frequencies above 10 kHz is due to the use of a band-pass filter to avoid interferences with the neighboring pixel.

## IV. DISCUSSION

Fig. 5 (*a*, *b* and *d*) shows a common trend regarding $\alpha$, $\beta$ and $\mathcal{L}$ for all the TESs with bars. Their values decrease at the beginning of the transition reaching a minimum around $0.25R_N$. Then they suddenly increase, forming a peak around the high part of the transition to end up eventually with very low values when the TES becomes normal. Such a peak has already been observed in other work [35,36] and looks consistent with the presence of metal bars. Moreover, a different alignment between these bars and the detector can induce a shifting of this peak [37]. Fig. 5 (*c*) illustrates faster time constants for TES9-TES10 compared to those for TES19-TES20, which is expected from the difference in their thermal conductance. From Fig. 6 we can conclude that the noise model used with the parameters obtained from the fitting of the complex impedance measurements is in good agreement with the experimental noise spectra. We also get quite low *M*-factor, as expected from detectors with bars [2].

On the other hand, we record different behavior from detectors without bars, i.e. higher values of $\alpha$, $\beta$ and *M*-factor. From Fig. 8 (*a*, *b* and *d*) we can conclude again that $\alpha$, $\beta$ and $\mathcal{L}$ have the same trend. They maintain generally higher values over the transition with a smother trend without any peak in the higher part of the transition compared to the detectors with bars. Also in this case we have two TESs with different thermal conductance values that are reflected in the faster response of the detector TES4 compared with the detector TES21 as pointed out in Fig. 8 (*c*). Fig. 9 shows that the detectors without bars have larger noise levels in the frequency band where the Johnson noise is dominant. This indicates that the $M$-factor is considerably higher here compared to that obtained from TESs with bars as reported in Tab.1 and consistent with those reported in [3,38].

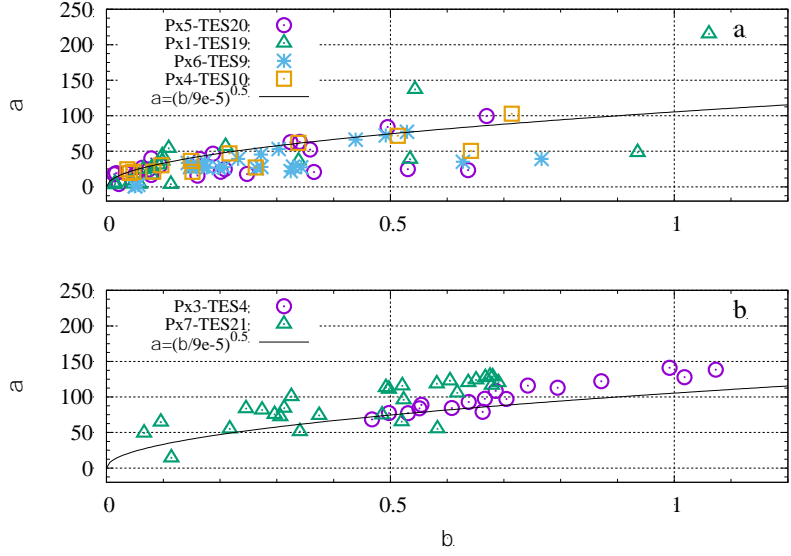

FIG. 10. α versus β for TESs with bars (a) and for TESs without bars (b). High α values for the TESs with bars have been intentionally left off-scale to make the comparison between the two type of detectors easier. The α/β ratio remains constant for TESs with the same size but with different absorber coupling or metal structures.

In fig. 10 we show a plot of $\alpha$ versus $\beta$ for TESs with bars (a) and for TESs without bars (b). As already remarked in [39], we do see similar values among the measured α/β ratio for pixel devices having the same size but with different stems for absorber coupling or presence of bars. This indicates a good uniformity and quality of our bilayer over the array. We also found that the correlation $\alpha = \sqrt{\beta/9 \times 10^{-5}}$ (line in Fig. 10) reported in [40] is still consistent with our experimental data.

TABLE I. Summary of the TES parameters as $\alpha$ and $\beta$ from the fitting of the complex impedance measurements and as $M$-factor obtained from the modeling of the noise spectra.

| TES | $\alpha$ | $\beta$ | M factor |
|---|---|---|---|
| TES20 (Bars) | $10 \leq \alpha \leq 150$ Peak @ 60-80% | $0 < \beta \leq 1$ Peak as α | $1 \leq M \leq 2$ |
| TES19 (Bars) | $10 \leq \alpha \leq 200$ Peak @ 70-90% | $0 < \beta \leq 1$ Peak as α | $1 \leq M \leq 2$ |
| TES10 (Bars) | $10 \leq \alpha \leq 250$ Peak @ 60-80% | $0 < \beta \leq 2$ Peak as α | $2 \leq M \leq 3$ |
| TES9 (Bars) | $10 \leq \alpha \leq 100$ Peak @ 70-90% | $0 < \beta \leq 1$ Peak as α | $2 \leq M \leq 3$ |
| TES4 (No Bars) | $80 \leq \alpha \leq 150$ Smooth, bigger in the middle of transition | $0.5 < \beta \leq 1.2$ Smooth over the bias points | $5 \leq M \leq 7$ |
| TES21 (No Bars) | $60 \leq \alpha \leq 140$ Smooth, bigger in the middle of transition | $0 < \beta \leq 0.8$ Smooth over the bias points | $5 \leq M \leq 7$ |

Table I summarizes the results in terms of $\alpha$, $\beta$ and M-factor for all the devices under test to highlight the main difference between TESs with and without bars that is eventually the main goal of this work.

## V. CONCLUSION

Complex impedance measurement is a well-known technique, which is widely used to study the performance of TESs under DC-bias. We have extended this technique to be performed in the AC-bias configuration that is intrinsic to the MHz FDM readout system. We have measured the complex impedance of different TESs located in a 5×5 array and obtained a good agreement between measurements and the fitting model. Good matching has also been reached between the measured noise spectra and the detector noise modeled using the parameters from the impedance fitting. In order to achieve a correct understanding of our detectors, statistical analysis of the residuals between the measured data and the fitting model have been discussed, demonstrating the goodness of the one-body thermal model compared to more complicated multi-body models.

By using this technique we have shown a complete set of $\alpha$ and $\beta$ values. The presence of metal structures on top of a TES does reduce the value of $\alpha$ and $\beta$, but can induce the appearance of localized peaks where their value increases significantly. On the other hand, TESs without bars show larger values of $\alpha$ and $\beta$, with a broad region of parameter space without peaks. The M-factor is demonstrated to be considerably higher in TESs without bars compared to those with bars. Despite this, the measured α/β ratio for pixel devices having the same size (although with different stem for absorber coupling or presence of bars) is comparable. This indicates a good uniformity and quality of our bilayer over the array.

Complex impedance measurement is not only a fundamental tool to get thermal and electrical parameters from a single detector but, we have confirmed that applied to an array, it can give a clear fingerprint of the different detectors under test.

**ACKNOWLEDGMENTS**

This work is partly funded by European Space Agency (ESA) and coordinated with other European efforts under ESA CTP contract ITT AO/1-7947/14/NL/BW. It has also received funding from the European Union's Horizon 2020 Programme under the AHEAD (Activities for the High-Energy Astrophysics Domain) project with grant agreement number 654215. The authors would like to thank Dr. Hiroki Akamatsu for his valuable advice on the statistical issues examined in this paper.